\documentclass[fleqn,usenatbib]{mnras}
\usepackage[T1]{fontenc}
\usepackage{ae,aecompl}
\usepackage{graphicx}
\usepackage{amsmath}
\usepackage{amssymb}
\usepackage{txfonts}
\usepackage{bm}

\graphicspath{./}

\newcommand{\ltsima}{$\; \buildrel < \over \sim \;$}
\newcommand{\lsim}{\lower.5ex\hbox{\ltsima}}
\newcommand{\gtsima}{$\; \buildrel > \over \sim \;$}
\newcommand{\gsim}{\lower.5ex\hbox{\gtsima}}

\newcommand{\dd}{\mathrm{d}}

\title[Cosmic Infrared Background anisotropies]
{The information content of Cosmic Infrared Background anisotropies}
\author[Reischke, Desjacques and Zaroubi]
{Robert Reischke\thanks{email:  \href{mailto:r.reischke@campus.technion.ac.il}{ r.reischke@campus.technion.ac.il}}$^{1,2}$, Vincent Desjacques$^1$ and Saleem Zaroubi$^{2,3,1}$
\\
$^1$ Department of Physics, Technion, Haifa 32000, Israel
\\
$^2$ Department of Natural Sciences, The Open University of Israel, 1 University Road, P.O. Box 808, Ra'anana 4353701, Israel
\\
$^3$ Kapteyn Astronomical Institute, University of Groningen, Landleven 12, 9747AD Groningen, the Netherlands
}

\begin{document}
\onecolumn

\pagerange{\pageref{firstpage}--\pageref{lastpage}}
\pubyear{2019}
\maketitle
\label{firstpage}

% --- abstract --- %
\begin{abstract}
We use analytic computations to predict the power spectrum as well as the bispectrum of Cosmic Infrared Background (CIB) anisotropies. Our approach is based on the halo model and takes into account the mean luminosity-mass relation. The model is used to forecast the possibility to simultaneously constrain cosmological, CIB and halo occupation distribution (HOD) parameters in the presence of foregrounds.  

For the analysis we use wavelengths in eight frequency channels between 200 and 900$\;\mathrm{GHz}$ with survey specifications given by \texttt{Planck} and \texttt{LiteBird}. We explore the sensitivity to the model parameters up to multipoles of $\ell =1000$ using auto- and cross-correlations between the different frequency bands. 
With this setting, cosmological, HOD and CIB parameters can be constrained to a few percent. Galactic dust is modeled by a power law and the shot noise contribution as a frequency dependent amplitude which are marginalized over.

We find that dust residuals in the CIB maps only marginally influence constraints on standard cosmological parameters. Furthermore, the bispectrum yields tighter constraints (by a factor four in $1\sigma$ errors) on almost all model parameters while the degeneracy directions are very similar to the ones of the power spectrum. The increase in sensitivity is most pronounced for the sum of the neutrino masses. Due to the similarity of degeneracies a combination of both analysis is not needed for most parameters. This, however, might be due to the simplified bias description generally adopted in such halo model approaches.
\end{abstract}

% --- keywords --- %

% --- section: introduction --- %
\section{Introduction}
A good fraction of the radiation emitted by stars in galaxies is absorbed by dust and re-emitted in the far infrared. The resulting diffuse background produced by distant galaxies is called the cosmic infrared background (CIB). 
Measurements of the CIB \citep[e.g.][]{dwek_cobe_1998,fixsen_spectrum_1998,planck_collaboration_planck_2011} therefore provide a window into the galaxy formation history of the Universe. In addition to its dependence on the star formation rate (out to fairly high redshifts), the CIB also furnishes a probe of the cosmological background as well as fluctuations in the galaxy distribution. Therefore, it also carries a wealth of cosmological information.

Anisotropies in the CIB have drawn a lot of attention since they have been detected e.g. by \citet{lagache_correlated_2007,viero_blast:_2009,hall_angular_2010,viero_measuring_2012} in the range $100\;\mu\mathrm{m}-1000 \;\mu\mathrm{m}$ or in the submilimeter. \citet{planck_collaboration_planck_2011} measured the CIB anisotropies with unprecedented accuracy, which since then has been updated and also cross-correlated with the lensing potential of the Cosmic Microwave Background (CMB) \citep{planck_collaboration_planck_2014}.
At the same time the theoretical modelling of the anisotropies underwent a lot of activities. While early models fitted biased linear power spectra \citep{lagache_correlated_2007,hall_angular_2010} these models have been replaced by more elaborate ones \citep{planck_collaboration_planck_2011}. Current models consist of two ingredients: a description of the evolution of the dark matter distribution and a model for the evolution of the galaxies which reside in the ambient dark matter as well as the connection between both. For the galaxies many models have been used which reproduce the observed differential number counts and luminosity functions \citep[e.g.][]{bethermin_modeling_2011}. The halo model \citep{cooray_halo_2002} in combination with the halo occupation distribution (HOD) offers a framework to study the spatial distribution of galaxies by linking the number of galaxies in a specific halo to its mass. This model was used by several authors to study the power spectrum and the bispectrum of the CIB \citep{viero_blast:_2009,penin_accurate_2012,lacasa_non-gaussianity_2014,penin_non-gaussianity_2014} and to put forecasts on constraints on the HOD parameters. \citet{shang_improved_2012} proposed an improved model to capture the mass dependence of the mean mass-luminosity relation in case of the CIB's power spectrum. Both models where used to fit the CIB measured by the \texttt{Planck} satellite \citep{planck_collaboration_planck_2014}. \citet{wu_minimal_2017} constructed an empirical model including stellar mass functions, the star-forming main sequence as well as dust attenquation. 

Owing to their extended redshift range, measurements of the far-CIB (like other intensity mappings) probe comoving volumes significantly larger than those accessible to forthcoming galaxy surveys (such as \texttt{Euclid} or LSST). On the other hand, they are mainly limited by contamination of the dust of the Milky Way. It is therefore necessary to either remove the galactic dust from the CIB maps or model it accordingly and marginalize over the dust component in the end. 
Thus, provided the foreground can either be removed or modelled, the anisotropies of the CIB can in principle be used to constrain HOD and cosmological parameters. \citet{tucci_cosmic_2016} showed for example how the power spectrum of the CIB anisotropies can potentially constrain local primordial non-Gaussianities at a level competitive with future galaxy surveys.

In this work we use the formalism developed in \citet{lacasa_non-gaussianity_2014} to extend the model described in \citet{shang_improved_2012} to the bispectrum. We only model the at least partially connected parts of the spectra using the halo model. The purely disconnected parts, i.e. the shot noise component, are treated as free parameters in the analysis to fit the angular spectra at high multipoles. We then study the impact of foregrounds given by galactic dust which we model by a power law for the spatial part and by a modified black-body spectrum for the frequency dependence. The impact of residual foregrounds is investigated for the power spectrum for a combined survey of the \texttt{Planck} and \texttt{LiteBird} frequencies above $200\;\mathrm{GHz}$ with a total of 8 frequency channels by studying the constraints on CIB, HOD and cosmological parameters. Including also the information of the bispectrum we then compare its performance with the power spectrum analysis and also give the joint constraints between both probes.
If not stated otherwise we will use the best fit CIB parameters from \citet{planck_collaboration_planck_2014} and the best fit cosmological parameters of \citet{planck_collaboration_planck_2018}.

The remainder of the paper is structured as follows: In \autoref{sec:model_CIB} we review the modelling of CIB anisotropies and give the explicit expressions up to the bispectrum. \autoref{sec:statistics} briefly introduces the statistical analysis. We show results in \autoref{sec:results} and conclude in \autoref{sec:conclusions}.

\section{CIB anisotropies}
In this section we briefly review the modelling of CIB anisotropies on the basis of the halo model and the HOD. We provide the equations for the angular power spectrum and the bispectrum. Furthermore, we will briefly discuss the shot noise component and galactic foregrounds. 
\label{sec:model_CIB}
\subsection{CIB anisotropies}
The specific infrared intensity at frequency $\nu$ is given by
\begin{equation}\label{eq:specific_intensity}
I_\nu = \int\mathrm{d}\chi a j_\nu (\chi(z)) =  \int\mathrm{d}\chi a \bar j_\nu (\chi(z))\left(1+\frac{\delta j_\nu(\chi(z))}{ \bar j_\nu (\chi(z))} \right)\;,
\end{equation}
where $j_\nu(\chi(z))$ is the specific emission coefficient and a bar indicates the average emissivity.  We integrate along the line-of-sight over the comoving distance:
\begin{equation}
    \chi(z(a)) = -c\int_1^a\frac{\mathrm{d}a}{a^2H(a)}\;,
\end{equation}
with the scale factor $a$ and the Hubble function $H\coloneqq \dot{a}/a$.
Introducing a spherical basis, $\delta I_\nu = \sum_{\ell,m}\delta I_{\ell m,\nu} Y_{\ell m} $, the correlation of the spherical harmonic coefficients defines the angular power spectrum:
\begin{equation}\label{eq:angular_power_spectrum_CIB}
\left\langle \delta I_{\ell m,\nu} \delta I_{\ell^\prime m^\prime, \nu^\prime} \right\rangle = C_{\ell,\nu\nu^\prime}\delta_{\ell\ell^\prime}\delta_{m m^\prime}\;,
\end{equation}
where the Kronecker deltas, $\delta_{\ell\ell^\prime}$ and $\delta_{mm^\prime}$, ensure spatial homogeneity and isotropy respectively.
Using the Limber approximation \citep{limber_analysis_1954}, the angular power spectrum can be calculated as:
\begin{equation}\label{eq:CIB_power_spec}
C_{\ell,\nu\nu^\prime} = \int\frac{\mathrm{d}\chi}{\chi^2}a^2\bar j_\nu (\chi(z))\bar j_{\nu^\prime} (\chi(z)) P_{j,\nu\nu^\prime}\left(\frac{\ell + 0.5}{\chi},\chi\right)\;,
\end{equation}
with the power spectrum of the emission coefficient:
\begin{equation}
(2\pi)^3 \bar j_\nu (\chi(z))\bar j_{\nu^\prime} (\chi(z))P_{j,\nu\nu^\prime}(k,\chi) \delta_\mathrm{D}^{(3)}(\bm k-\bm k^\prime) =\langle\delta j_\nu(\bm k)\delta j_{\nu^\prime}(\bm k^\prime) \rangle \; .
\end{equation}
One can now equate $P_{j,\nu\nu^\prime}$ with the power spectrum of galaxies. This assumes that spatial variations in the emission coefficient are sourced by galaxies, such that $\delta  j_\nu/ \bar j _\nu = \delta n_\mathrm{gal}/\bar{n}_\mathrm{gal}$ and that there are no other biases apart from the galaxy bias itself. The above procedure generalises to higher order spectra. For the angular bispectrum defined by:
\begin{equation}
\label{eq:bispectrum}
\left\langle\delta I_{\ell_1 m_1,\nu_1}\delta I_{\ell_2 m_2,\nu_2}\delta I_{\ell_3 m_3,\nu_3} \right\rangle = 
\begin{pmatrix}
\ell_1 & \ell_2 & \ell_3 \\
m_1 & m_2 & m_3 
\end{pmatrix}
B_{\nu_1,\nu_2,\nu_3}(\ell_1,\ell_2,\ell_3)\;,
\end{equation}
where the Wigner 3$j$ symbol was introduced. $B_{\nu_1,\nu_2,\nu_3}(\ell_1,\ell_2,\ell_3)$ can be expressed as follows, again using the Limber approximation
\begin{equation}\label{eq:CIB_bispectrum}
B_{\nu_1,\nu_2,\nu_3}(\ell_1,\ell_2,\ell_3) = \int\frac{\dd \chi}{\chi^4}\bar{j}_{\nu_1}(\chi)\bar{j}_{\nu_2}(\chi)\bar{j}_{\nu_3}(\chi) a^3(\chi)B_{j,\nu_1\nu_2\nu_3}\left(\frac{\ell_1 + 0.5}{\chi},\frac{\ell_2 + 0.5}{\chi},\frac{\ell_3 + 0.5}{\chi},\chi\right) \;,
\end{equation}
with the bispectrum of the emissivity coefficient $B_{j,\nu_1\nu_2\nu_3}(k_1,k_2,k_3,\chi)$, given by 
\begin{equation}
(2\pi)^3 \bar j_{\nu_1} (\chi(z))\bar j_{\nu_2} (\chi(z))\bar j_{\nu_3}(\chi(z))B_{j,\nu_1\nu_2\nu_3}(k_1,k_2,k_3,\chi) \delta_\mathrm{D}^{(3)}(\bm k_{123}) =\langle\delta j_{\nu_1}(\bm k_1)\delta j_{\nu_2}(\bm k_2)\delta j_{\nu_3}(\bm k_3) \rangle \; ,
\end{equation}
where $\delta_\mathrm{D}^{(3)}(\bm k_{123})$ ensures that the three wave vectors form a proper triangle. As for the power spectrum we will relate the bispectrum of the emissivity coefficient to the galaxy bispectrum (\ref{eq:galaxy_bispectrum_hm}).

\subsection{Halo model}
The connection between galaxies and dark matter can be described using the halo model together with the HOD. The galaxy power spectrum is generally given by:
\begin{equation}
P_\mathrm{gal}(k,z) = P_{1\mathrm{h}}(k,z) + P_{2\mathrm{h}}(k,z) + P_\mathrm{shot}(k,z)\; ,
\end{equation}
with the $1-$halo $2-$halo and shot noise term respectively:
\begin{equation}
\begin{split}
P_\mathrm{1h} = & \int \mathrm{d}M \frac{\dd n}{\dd M}(M,z) \frac{2N_\mathrm{cen}(M,z) N_\mathrm{sat}(M,z) + N^2_\mathrm{sat}(M,z)}{\bar{n}^2_\mathrm{gal}}u^2(k|M,z) \;,\\
P_\mathrm{2h} = & \left(\int \mathrm{d}M \frac{\dd n}{\dd M}(M,z) \frac{N_\mathrm{cen}(M,z) + N_\mathrm{sat}(M,z)}{\bar{n}_\mathrm{gal}}b_1(M,z)u(k|M,z)\right)^2P_\mathrm{lin}(k,z)\;, \\
P_\mathrm{shot} = & \frac{1}{\bar{n}_\mathrm{gal}}\;.
\end{split}
\end{equation}
Here, $\dd n/\dd M$ is the halo mass function for which we use the \citep{tinker_toward_2008} fitting formula. $u(k|M,z)$ the Fourier transform of the density profile of a halo at given mass and redshift:
The density profile of the halos dictates the small scale clustering properties of the galaxies. For a NFW halo \citep{navarro_universal_1997}, the Fourier transform of the density profile is given by
\begin{equation}
u(k|M,z) = \cos(kr_\mathrm{s})\left[\mathrm{Ci}(k(1+c)r_\mathrm{s}) - \mathrm{Ci}(kr_\mathrm{s})\right] -\frac{\sin(ckr_\mathrm{s})}{kr_\mathrm{s}(1+c)} + \frac{\sin(kr_\mathrm{s})\left(\mathrm{Si}(kr_\mathrm{s}(1+c))-\mathrm{Si}(kr_\mathrm{s})\right)}{\frac{1}{1+c} +\log(1+c) -1}\;.
\end{equation}
The concentration $c$ is given by an empirical relation and the scaling radius is given by
\begin{equation}
r_\mathrm{s} = \frac{r_\mathrm{vir}}{c} = \left(\frac{3M}{4\pi \Delta_V \bar\rho_\mathrm{m}c^3}\right)^{1/3}\;,
\end{equation}
with $\Delta_V = 200$ and $\bar\rho_\mathrm{m}$ the average matter density.  $\bar{n}_\mathrm{gal}$ is the mean number density of galaxies defined as
\begin{equation}
\bar{n}_\mathrm{gal}(z) = \int\dd M [N_\mathrm{sat}(M,z) + N_\mathrm{cen}(M,z)]\frac{\dd n}{\dd M}(M,z)\;.
\end{equation}
In this expression, $N_\mathrm{gal}(M,z) = N_\mathrm{sat}(M,z) + N_\mathrm{cen}(M,z)$ is the average number of galaxies in halos of mass $M$ at redshift $z$. $N_\mathrm{cen}$ and $N_\mathrm{sat}$ denotes the contribution from central and satellite galaxies, respectively. 
HODs suggest that the average number of satellite and central galaxies can be parametrised as follows:
\begin{equation}\label{eq:satellites_and_central}
N_\mathrm{sat}(M,z) = \frac{1}{2}\left[1+\mathrm{erf}\left(\frac{\log_{10}(M)-\log_{10}(2M_\mathrm{min}) }{\sigma_{\log_{10} M}}\right)\right]\left(\frac{M}{M_\mathrm{sat}}\right)^{\alpha_\mathrm{sat}}, \ \ \ \ N_\mathrm{cen}(M,z) = \frac{1}{2}\left[1+\mathrm{erf}\left(\frac{\log_{10}(M)-\log_{10}(M_\mathrm{min}) }{\sigma_{\log_{10} M}}\right)\right]\; ,
\end{equation}
in which $M_\mathrm{min}$ and $\alpha_\mathrm{sat}$ are determined by observations.
In our model, we use the sub-halo mass function, ${\dd N_\mathrm{sub}}/{\dd m}$, which we take to be of the form \citep{tinker_what_2010}
\begin{equation}\label{eq:sub_halo_mf}
\frac{\dd N_\mathrm{sub}}{\dd \log m} (m|M) =0.3\left(\frac{m}{M}\right)^{-0.7}\exp\left[-9.9\left(\frac{m}{M}\right)^{2.5}\right]\,,
\end{equation}
where $M$ and $m$ are the mass of the parent halo and the subhalo, respectively. The number of satellite galaxies can then be computed as:
\begin{equation}
    N_\mathrm{sat}(M,z) = \int\dd\log m \frac{\dd N_\mathrm{sub}}{\dd \log m} (m|M)\;.
\end{equation}
Finally, the first order bias, $b_1(\nu)$, is chosen such that the constraint \citep{tinker_toward_2008}
\begin{equation}\label{eq:1st_order_bias}
1 = \int \mathrm{d}\nu b_1(\nu)f(\nu)\;,
\end{equation}
is fulfilled subject to the constraint of the mass function. 
In similar fashion one finds for the bispectrum of galaxies in the halo model \citep[e.g.][]{lacasa_non-gaussianity_2014} the following relations:
\begin{equation}\label{eq:galaxy_bispectrum_hm}
B_\mathrm{gal}(k_1,k_2,k_3,z) = B_{1h}(k_1,k_2,k_3,z) + B_{2h}(k_1,k_2,k_3,z) + B_{3h}(k_1,k_2,k_3,z) + B_{\mathrm{shot}1h}(k_1,k_2,k_3,z) + B_{\mathrm{shot}2h}(k_1,k_2,k_3,z)\;,
\end{equation}
where
\begin{equation}
\begin{split}
\label{eq:galaxy_bispectrum_contributions}
B_{1h}(k_1,k_2,k_3,z) = & \ \int\dd M\frac{\dd n}{\dd M}(M,z) u(k_1|M,z)u(k_2|M,z)u(k_3|M,z)\frac{N_\mathrm{sat}^3(M,z) + 3N_\mathrm{cen}(M,z)N_\mathrm{sat}^2(M,z)}{\bar{n}_\mathrm{gal}(z)}\; ,\\
B_{2h}(k_1,k_2,k_3,z) = & \ \ \mathcal{G}_1(k_1,k_2,z)P_\mathrm{lin}(k_3,z)\mathcal{F}_1(k_3,z) + \mathcal{G}_1(k_1,k_3,z)P_\mathrm{lin}(k_2,z)\mathcal{F}_1(k_2,z) + \mathcal{G}_1(k_2,k_3,z)P_\mathrm{lin}(k_1,z)\mathcal{F}_1(k_1,z) \; , \\
B_{3h}(k_1,k_2,k_3,z) = & \ \ \mathcal{F}_1(k_1,z) \mathcal{F}_1(k_2,z)\mathcal{F}_1(k_3,z)\left[F_2(\bm{k_1},\bm{k_2}) P_\mathrm{lin}(k_1,z)P_\mathrm{lin}(k_2,z) + \mathrm{perm} \right] \\ &  + \ \mathcal{F}_1(k_1,z) \mathcal{F}_1(k_2,z)\mathcal{F}_2(k_3,z) P_\mathrm{lin}(k_1,z)P_\mathrm{lin}(k_2,z) + \mathrm{perm} \;.
\end{split}
\end{equation}
The following functions have been defined for shorthand convenience:
\begin{equation}\label{eq:bispectrum_functions}
\begin{split}
\mathcal{F}_1(k,z) &  =\  \int\dd M\frac{N_\mathrm{gal}(M,z)}{\bar{n}_\mathrm{gal}(z)}\frac{\dd n}{\dd M} (M,z)b_1(M,z)u(k|M,z)\; , \\
\mathcal{F}_2(k,z) &  =\  \int\dd M\frac{N_\mathrm{gal}(M,z)}{\bar{n}_\mathrm{gal}(z)}\frac{\dd n}{\dd M} (M,z)b_2(M,z)u(k|M,z)\; , \\
\mathcal{G}_1(k_1,k_2,z) &  =\  \int\dd M\frac{2N_\mathrm{cen}(M,z) N_\mathrm{sat}(M,z) + N^2_\mathrm{sat}(M,z)}{\bar{n}^2_\mathrm{gal}}\frac{\dd n}{\dd M} (M,z)b_1(M,z)u(k_1|M,z)u(k_2|M,z)\; , \\
F_2(\bm{k_1},\bm{k_2}) & = \ \frac{5}{7}+\frac{1}{2}\cos(\theta_{12})\left(\frac{k_1}{k2} + \frac{k_2}{k_1}\right) +\frac{2}{7}\cos^2(\theta_{12})\; ,
\end{split}
\end{equation}
with the first and second order bias $b_1$ and $b_2$ respectively. The second order bias is given by a fitting equation given by \citep{lazeyras_precision_2016}:
\begin{equation}\label{eq:2nd_order_bias}
b_2(\nu) = 0.412 - 2.143b_1 + 0.929b_1^2 +0.008b_1^3\;.
\end{equation}
The quantity $\nu$ describes the peak-background split threshold with $\nu = \delta_\mathrm{c}/\sigma(M,z)$. For the mass function we take the \citep{tinker_toward_2008} and a consistent expression for the linear bias $b_1$. $\sigma(M,z)$ is the standard deviation of the density field smoothed at a mass scale 
\begin{equation}
\label{eq:variance_scale}
\sigma^2(M) = \frac{1}{2\pi^2}\int k^2\dd k \left(\frac{j_1(kR(M))}{kR(M)}\right)^2 P_\mathrm{lin}(k) \;,
\end{equation}
with the spherical Bessel function $j_1$ and 
\begin{equation}\label{eq:radius_mass}
R^3(M) = \frac{3M}{4\pi \bar\rho_\mathrm{m}(a)}\;.
\end{equation}
The linear power spectrum is calculated with \textsc{Class} \citep{lesgourgues_cosmic_2011} assuming the fiducial cosmology outlined above.

\subsection{Mean emissivity}
Following \citet{shang_improved_2012}, we will sum up all galaxies contributing to the CIB luminosity at a given frequency and redshift weighted by their differential number density, i.e. the halo mass function:
\begin{equation}\label{eq:mean_emissivity}
\bar j_\nu(z) = \int\dd M\frac{\dd n}{\dd M}\left[f_\nu^c(M,z)+f_\nu^s(M,z)\right]\;.
\end{equation}
In particular we split the mean emissivity into a contribution from central, $f_\nu^c(M,z)$, and satellite galaxies, $f_\nu^s(M,z)$ which can be calculated by:
\begin{equation}\label{eq:contribution_central_satellites}
f_\nu^c(M,z) = \frac{1}{4\pi}N_c(M,z)L_{c,(1+z)\nu}(M,z), \quad f_\nu^s(M,z) = \frac{1}{4\pi}\int_0^M\dd m \frac{\dd N_\mathrm{sub}}{\dd m}(m,z|M)L_{s,(1+z)\nu}(m,z)\;.
\end{equation}
This definition ensures that the contribution to the total emissivity depends on the mass not only by the number density of sources at a given mass, but also on the mass-luminosity relation, which is encoded in $L_{(1+z)\nu}(M,z)$.
For simplicity, the infrared luminosity is assumed to be the same for central and satellite galaxies:
\begin{equation}\label{eq:luminosity_mass_relation}
L_{(1+z)\nu}(M,z) = L_0(1+z)^\delta\frac{M}{\sqrt{2\pi\sigma^2_{L/M}}}\exp\left[-\frac{\log(M)-\log(M_\mathrm{eff})}{2\sigma_{L/M}^2}\right]\Theta_\mathrm{CIB}[(1+z)\nu]\;.
\end{equation}
The luminosity peaks at a halo mass $M_\mathrm{eff}$ around which the negative feedback from supernovae and AGN on the star formation rate is minimum.
The overall amplitude $L_0$ is determined by fits to data. We will discuss this point further in section \ref{sec:results}. For the spectral energy distribution (SED) of the galaxies we assume a modified black-body spectrum with a power law emissivity
\begin{equation}\label{eq:galaxy_sed}
\Theta_\mathrm{CIB} =\left\{ 
\begin{array}{cc}
\left(\frac{\nu}{\nu_0}\right)^\beta \frac{B_\nu(T_\mathrm{d})}{B_{\nu_0}(T_\mathrm{d})} & \nu\leq\nu_0 \;, \\
\left(\frac{\nu }{\nu_0}\right)^\gamma & \nu > \nu_0\; .
\end{array}\right.
\end{equation}
Here $B_\nu$ is the Planck function and $T_\mathrm{d}$ is the dust temperature for which we assume the following redshift dependence:
\begin{equation}\label{eq:dust_temperature}
T(z) = T_0(1+z)^\alpha\;.
\end{equation}
The two regimes are smoothly connected at $\nu_0$ such that
\begin{equation}\label{eq:connection}
\frac{\mathrm{\dd \log\Theta}}{{\dd\log\nu}}=-\gamma\;.
\end{equation}
Replacing the relations (\ref{eq:satellites_and_central}) by (\ref{eq:contribution_central_satellites}) fully specifies the model and takes into account the mass dependence of the luminosity as described in \citet{shang_improved_2012}. The fiducial parameters of our model are summarized in \autoref{table:cib_parameters}. 

\begin{table}
\begin{center}
\begin{tabular}{ccc}
parameter & fiducial value & description \\ \hline\hline
$\alpha$ & 0.36 & exponent of the dust temperature's redshift dependence \\
$T_0$ & $24.4\;\mathrm{K}$ & dust temperature today \\
$\beta$ & 1.75 & modification to the CIB's blackbody spectrum \\
$\gamma$ & 1.7 & CIB's power law emissivity at high frequencies \\
$\delta$ & 3.6 & exponent of the CIB's normalization redshift evolution \\
$M_\mathrm{eff}$ & $10^{12.6}\; M_\odot$ & peak of the specific CIB emissivity \\
$\sigma_{L/M}^2$ & 0.5 & range of halo masses producing a certain emissivity \\
$M_\mathrm{c}$ & $3\times 10^{11} M_\odot/h$ & minimum mass for a halo to host a central galaxy
\end{tabular}
\caption{Best fit CIB and halo occupation parameters together with their description. The parameters are chosen to fit the power spectra measurements of the data from \citep{planck_collaboration_planck_2014}.}
\label{table:cib_parameters}
\end{center}
\end{table}

.

\subsection{Shot noise}
\begin{figure}
\begin{center}
\includegraphics[width = 0.9\textwidth]{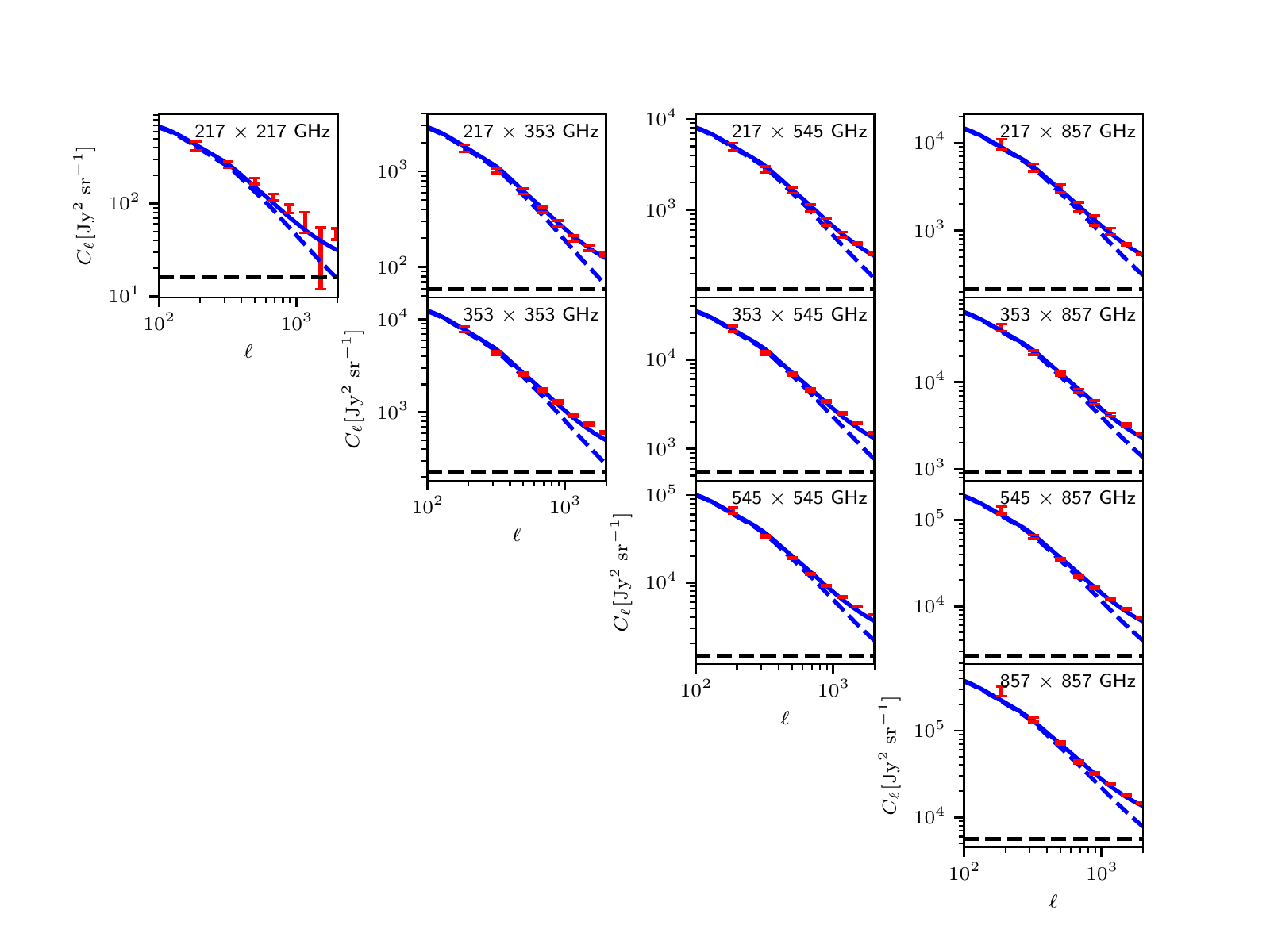}
\caption{CIB anisotropy angular power spectra as measured by \citet{planck_collaboration_planck_2014} in four frequency bands. The data points are shown in red. The dashed blue line corresponds to the clustering contribution while the dashed black line shows the shot noise contribution in each band. In solid blue we show the sum of clustering and shot noise.}
\label{fig:Planck_spectra}
\end{center}
\end{figure}

\begin{figure}
\begin{center}
\includegraphics[width = 0.5\textwidth]{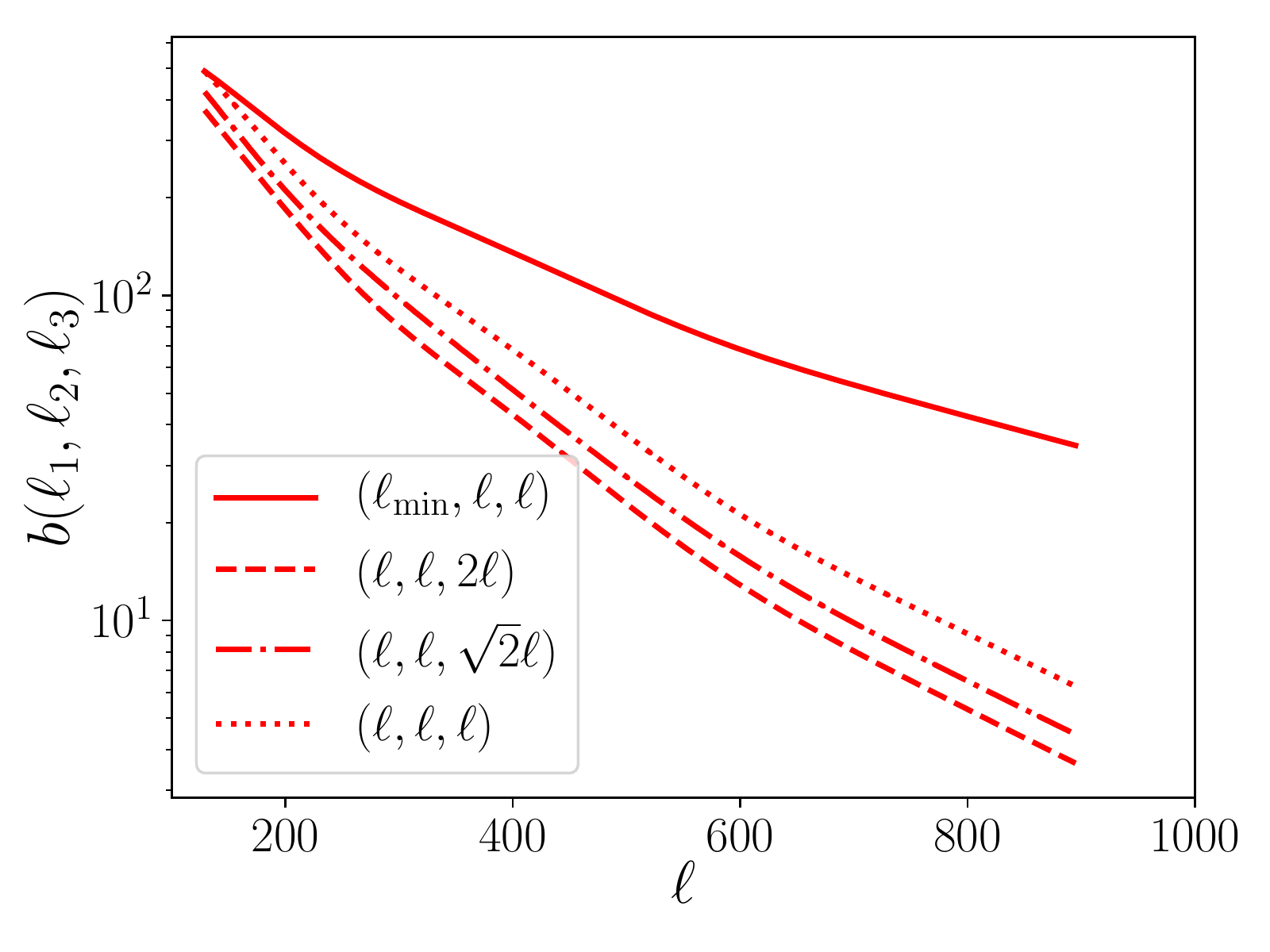}
\caption{Bispectrum for $\nu_1=\nu_2=\nu_3 = 353\;\mathrm{GHz}$ for different triangle configurations without shot noise contribution, that is $b(\ell_1,\ell_2,\ell_3) = b_{1h}+b_{2h} + b_{3h}$. Clearly the squeezed limit is the dominant contribution for the multipole range shown in the figure.}
\label{fig:Planck_bispectra}
\end{center}
\end{figure}

%\vd{As we already discussed, shot noise and 1-halo term basically mean the same thing here, isn't it ?}
As discussed in \citet{shang_improved_2012}, the model could in principle describe the shot noise term (which could in principle absorb the constant low-$\ell$ piece of the 1-halo term) originating from local fluctuations in the number density of galaxies. However, since this shot-noise is mainly sourced by the scatter in the luminosity-mass relation, which is not included in expression (\ref{eq:luminosity_mass_relation}), it will generally be underestimated by the model. In principle, there exist parametric models for the shot noise \citep{bethermin_modeling_2011}. Notwithstanding, we will remain agnostic and treat the shot noise amplitude as a free parameter like, for example, in the analysis performed in \cite{planck_collaboration_planck_2014}.
In \autoref{fig:Planck_spectra}, we show the CIB spectra as calculated for a Planck cosmology with the parameters from \autoref{table:cib_parameters}. The dashed blue line shows the clustering contribution, i.e. the sum of the one- and the two-halo term. In black the shot noise level is shown, while the solid blue curves shows the sum of all contributions. Furthermore, \autoref{fig:Planck_bispectra} shows the bispectrum at $353\;\mathrm{GHz}$ for different triangular configurations including only the clustering terms.

\subsection{Galactic dust}
The main foreground at infrared frequencies exceeding 200$\;\mathrm{GHz}$ is the Galactic dust emission. Like the SED of the CIB, its frequency spectrum is also well described by a modified black-body spectrum of the form \citep{planck_collaboration_planck_2014}:
\begin{equation}\label{eq:sed_dust}
\Theta_\mathrm{d}(\nu) = \left(\frac{\nu}{\nu_0}\right)^{\beta_\mathrm{d}}\frac{B_\nu(T_\mathrm{d})}{B_{\nu_0}(T_\mathrm{d})}\;.
\end{equation}
The reference frequency $\nu_0$ is chosen to be $353\;\mathrm{GHz}$. The best-fit dust temperature (assumed to be constant for all galaxies) and spectral index are $T_\mathrm{d}=19.6\;\mathrm{K}$, $\beta_\mathrm{d}=1.53$. We split the dust power spectra into a spatial correlation and a frequency correlation 
\begin{equation}
C^\mathrm{d}_{\ell,\nu\nu^\prime} =C_\ell^\mathrm{d}(\nu_0)\Theta_\mathrm{d}(\nu)\Theta_\mathrm{d}(\nu^\prime)R_{\nu\nu^\prime}\;,
\end{equation}
where the frequency correlation matrix $\boldsymbol{R}$ is given by \citep{tegmark_removing_1998}
\begin{equation}
\label{eq:frequency_correlation}
R_{\nu\nu^\prime} = \exp\left[-\frac{1}{2}\left(\frac{\log(\nu/\nu^\prime)}{\zeta}\right)^2\right]\;.
\end{equation}
The frequency coherence $\zeta$ encodes the strength of the correlation such that for $\zeta\to 0$, $\boldsymbol{R}\to\boldsymbol{\mathrm{id}}$, i.e. correlations between different frequency channels are absent. Conversely, $\zeta\to\infty$ corresponds to maximally correlated channels.

Since the contamination from Galactic dust is most severe in the Galactic plane, the amplitude of the dust power spectrum strongly depends on the sky fraction, $f_\mathrm{sky}$ (of the least contaminated pixels) considered. Following \cite{collaboration_planck_2015,miville-deschenes_statistical_2007}, we assume
\begin{equation}\label{eq:dust_power_amplitude}
C_\ell^\mathrm{d}(\nu_0)=1.45\times 10^6\left(\frac{f_\mathrm{sky}}{0.6}\right)^{4.6 +7.11\log(f_\mathrm{sky}/0.6)}\ell^{\alpha_\mathrm{d}}\;,
\end{equation}
where $\alpha_\mathrm{d}$ describes the spatial clustering of the foreground dust. Clearly, a lower frequency coherence will reduce the constraints on these parameters significantly. However, note that, in \cite{collaboration_planck_2015}, the dispersion of the dust emissivity index was measured to be 0.07. This corresponds to $\zeta = 10.1$, which yields an almost perfect correlation over the range of frequencies considered here.

\begin{table}
\begin{center}
\begin{tabular}{cccc}
band $\nu\;[\mathrm{GHz}]$ & $w_\nu\;[\mathrm{Jy}\;\mathrm{sr}^{-1}]$ & $ \theta_\mathrm{FWHM}\;[\mathrm{arcmin}]$ & experiment
\\ \hline\hline
217 & 43.32 & 5.02 & \texttt{Planck}\\
353 & 164.7 & 4.94 & \\
545 & 185.3 & 4.83 & \\
857 & 157.9 & 4.64  & \\ \hline
235 & 0.36 & 30.0 & \texttt{LiteBird}\\
280 & 1.45 & 30.0 & \\
337 &1.1 & 30.0 &  \\
402 &0.7 & 30.0 &  \\
\end{tabular}
\caption{Frequency bands and the corresponding white noise level $w_\nu$ and the angular resolution induced by the beam width $\theta_\mathrm{FWHM}$ for five frequency bands of \texttt{Planck} and the frequency bands of \texttt{LiteBird} above 200$\;\mathrm{GHz}$.}
\label{table:experiments}
\end{center}
\end{table}

\section{Statistics}
\label{sec:statistics}

\subsection{Fluctuations in spherical harmonics}
We consider $n_\mathrm{bands}$ maps of CIB intensities, $\delta I_\nu$, $\nu = \nu_1,...,\nu_{n_\mathrm{bands}}$, at frequency $\nu$ decomposed into in spherical harmonic coefficients $\delta I_{\ell m,\nu}$:
\begin{equation}\label{eq:spherical_harmonic_expansion}
\delta I_\nu (\bm{\hat{n}}) = \sum_{\ell,m} \delta I_{\ell m,\nu} Y_{\ell m} (\bm{\hat{n}})\;,
\end{equation} 
where $\bm{\hat{n}}$ is the direction of the line of sight and $Y_{\ell m}$ are the spherical harmonics.
Eq. (\ref{eq:CIB_power_spec}) describes the correlation of these modes which have to be diagonal in $m$ and $\ell$ due to, respectively, the statistical isotropy and homogeneity of the fluctuations. 

\begin{figure}
\begin{center}
\includegraphics[width = 0.45\textwidth]{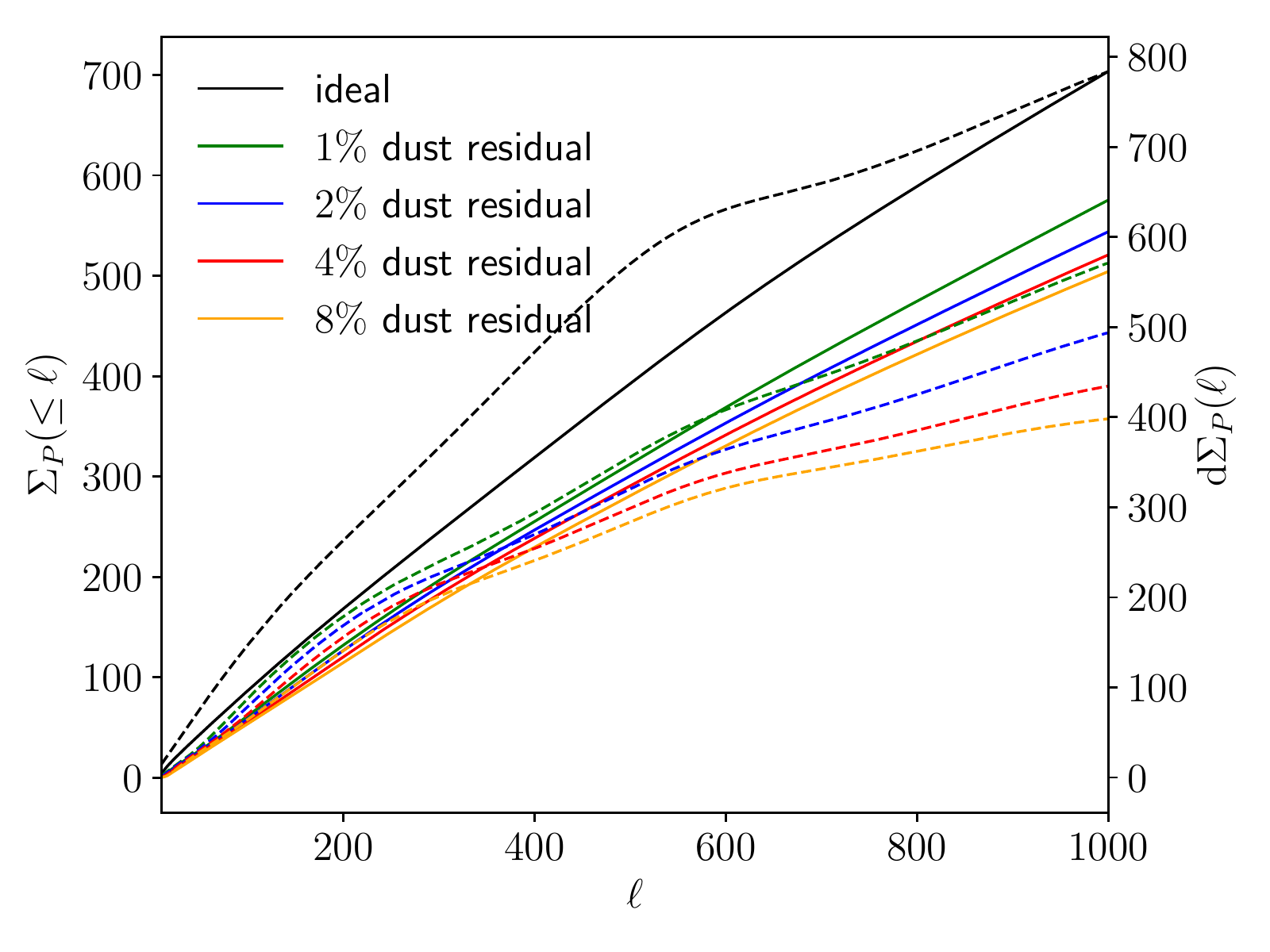}
\includegraphics[width = 0.45\textwidth]{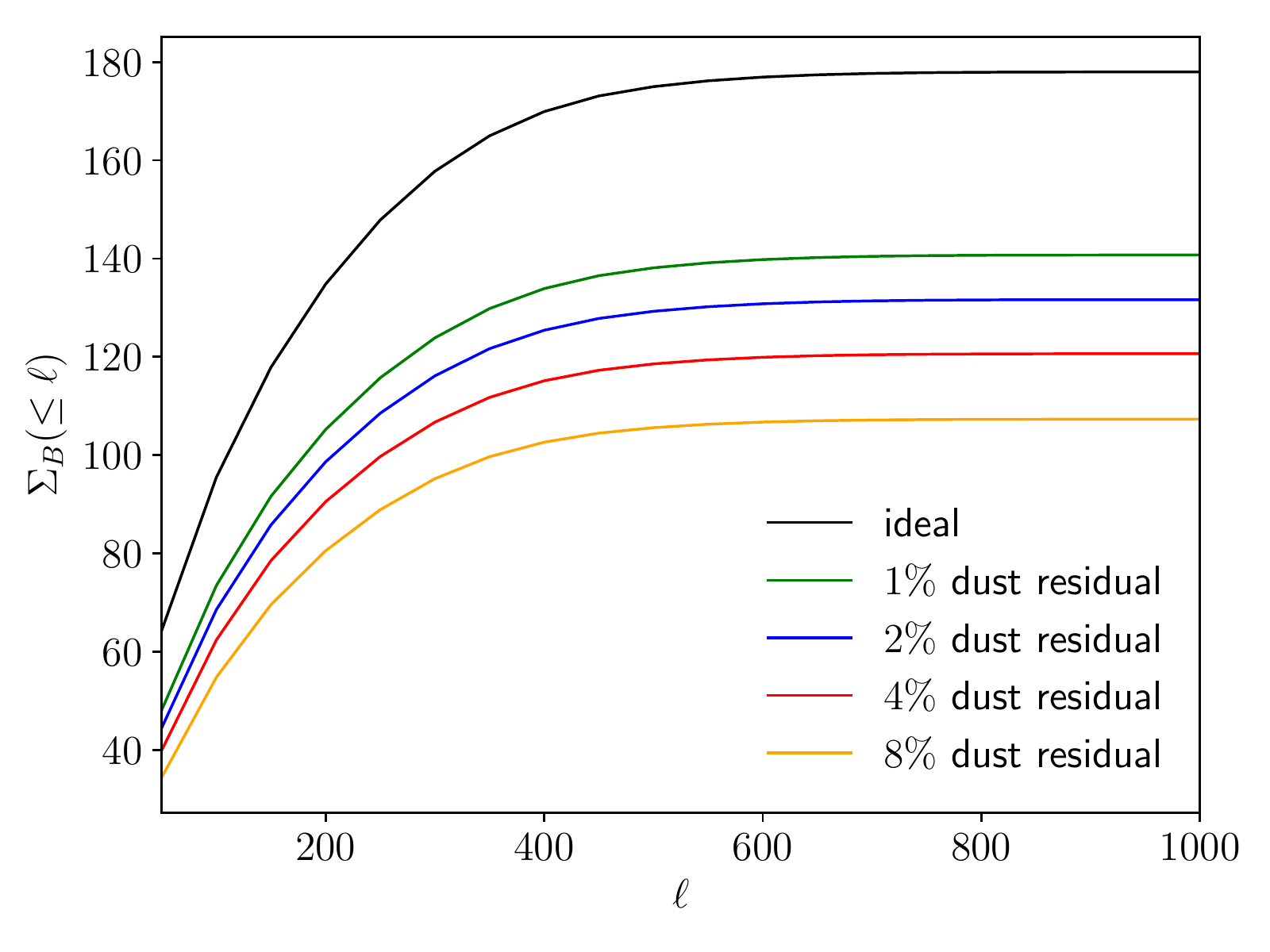}
\caption{Signal-to-noise ratio (SNR) for CIB maps whose noise is given only by instrumental noise, cosmic variance and possibly residual dust. \textit{Left}: Cumulative SNR of a power spectrum as solid lines, while dashed lines correspond to the differential SNR. \textit{Right}: Cumulative SNR for a bispectrum measurement.}
\label{fig:SNR_ideal}
\end{center}
\end{figure}

\subsection{Power spectrum}
For Gaussian fields we can express the probability of finding a set of modes $\{\delta I_{\ell m,\nu}\}$ given a model $\bm{\theta}$ by
\begin{equation}
p\left(\left\{\delta I^{\mathstrut}_{\ell m,\nu}\right\}\big|\bm{\theta}\right) \propto \prod_\ell \left(\mathrm{det}\left(\bm{C}_\ell^{-1}\right) \exp\left[\bm{\delta I}_{\ell m}^\dagger \boldsymbol{C}_\ell^{\mathstrut -1}\bm{\delta I}^{\mathstrut}_{\ell m}\right]\right)^{2\ell+1}\;,
\end{equation}
where we bundled all maps into a vector $\bm{\delta I}_{\ell m}$. Their covariance is given by $\boldsymbol{C}_\ell = \left\langle \bm{\delta I}^{\mathstrut}_{\ell m}\bm{\delta I}_{\ell m}^\dagger \right\rangle$, where the average is applied over all possible realizations of the data. The entries of the covariance are thus given by Eq. (\ref{eq:CIB_power_spec}). The observed spectra, $\hat{C}_{\ell,\nu\nu^\prime}$ include instrumental noise terms which are given by
\begin{equation}\label{eq:observed_spectra}
\hat{C}^{\mathstrut}_{\ell,\nu\nu^\prime} = C^{\mathstrut}_{\ell,\nu\nu^\prime} + N^{\mathstrut}_\ell(\nu)\delta^{\mathrm{K}}_{\nu\nu^\prime}\;,\quad N_\ell(\nu) = w_\nu\exp\left[\ell(\ell+1)\frac{\theta^2_{\mathrm{FWHM}}(\nu)}{8\log 2}\right]\;.
\end{equation}
Here,$w_\nu$ describes the instrumental white noise and $\theta^2_{\mathrm{FWHM}}$ the beam's width. We summarize the experimental settings in \autoref{table:experiments}.
The signal-to-noise ratio (SNR) for this setup is now readily computed as
\begin{equation}\label{eq:SNR_power_spec}
\Sigma^2(\leq\ell_\mathrm{max}) = f_\mathrm{sky}\sum_{\ell = \ell_\mathrm{min}}^{\ell_\mathrm{max}} \frac{2\ell +1}{2} \mathrm{tr}\left(\bm{C}^{\mathstrut}_\ell\bm{\hat{C}}_\ell^{{\mathstrut}-1}\bm{C}^{\mathstrut}_\ell\bm{\hat{C}}_\ell^{{\mathstrut}-1}\right) \;,
\end{equation}
where $f_\mathrm{sky}$ is the sky fraction compensating for incomplete sky-coverage. Likewise, the Fisher matrix is given by \citep{tegmark_karhunen-loeve_1997}:
\begin{equation}\label{eq:fisher_matrix_gaussian_power}
F_{ij}(\bm{\theta}_0) = f_\mathrm{sky}\sum_{\ell = \ell_\mathrm{min}}^{\ell_\mathrm{max}} \frac{2\ell +1}{2} \mathrm{tr}\left(\bm{\hat{C}}_\ell^{{\mathstrut}-1}\partial^{\mathstrut}_i\bm{C}^{\mathstrut}_\ell \bm{\hat{C}}_\ell^{{\mathstrut}-1}\partial^{\mathstrut}_j\bm{C}^{\mathstrut}_\ell\right)\bigg|_{\bm{\theta} = \bm{\theta}_0} \;,
\end{equation}
where $\partial_i$ is the derivative with respect to the $i$-th model parameter. 

\subsection{Bispectrum}
The spherical harmonic bispectrum in Eq. (\ref{eq:bispectrum}) is related to the flat sky bispectrum, $B_{\nu_1,\nu_2,\nu_3}(\bm{\ell}_1,\bm{\ell}_2,\bm{\ell}_3)$, through the relation
\begin{equation}\label{eq:flat_sky_bispectrum}
B_{\nu_1,\nu_2,\nu_3}(\ell_1,\ell_2,\ell_3) \simeq 
\begin{pmatrix}
\ell_1 & \ell_2 & \ell_3 \\
0 & 0 & 0 
\end{pmatrix}
\left(\frac{(2\ell_1+1)(2\ell_2+1)(2\ell_3+1)}{4\pi}\right)^{1/2}B_{\nu_1,\nu_2,\nu_3}(\bm{\ell}_1,\bm{\ell}_2,\bm{\ell}_3)\;,
\end{equation}
i.e. $\bm{\ell}_i$ are two dimensional vectors on the flat sky. It is thus consistent with the calculation of the bispectrum, Eq. (\ref{eq:CIB_bispectrum}) which indeed uses the flat sky approximation. The Wigner $3j$ symbol arises with $m_1 = m_2 = m_3 = 0$ originates from the integration over the Legendre polynomials, ensuring that the triangular inequality is satisfied.
Eq. (\ref{eq:CIB_bispectrum}) provides an explicit expression for the flat-sky bispectrum. Statistical homogeneity is ensured by the fact that the three multipole vectors must form a triangle. We assume a Gaussian covariance for the bispectrum, thus ignoring connected contributions from $n>2$ correlators. Furthermore we enforce the condition $\ell_1\leq \ell_2\leq\ell_3$ so that each triangle configuration is only counted once. With these approximations, the covariance of the bispectrum takes the simple form
\begin{equation}\label{eq:covariance_bispectrum}
\mathrm{Cov}\left[B_{{\mathstrut}\nu^{\mathstrut}_1\nu^{\mathstrut}_2\nu^{\mathstrut}_3}(\ell_1,\ell_2,\ell_3) B_{{\mathstrut}\nu^{{\mathstrut}\prime}_1\nu^{{\mathstrut}\prime}_2\nu^{{\mathstrut}\prime}_3}(\ell^{{\mathstrut}\prime}_1,\ell^\prime_2,\ell^\prime_3) \right] = \Delta(\ell_1,\ell_2,\ell_3) f_\mathrm{sky}^{-1} \left[\hat{C}_{\ell^{\mathstrut}_1,\nu^{\mathstrut}_1\nu_1^\prime}\hat{C}_{\ell^{\mathstrut}_2,\nu^{\mathstrut}_2\nu_2^\prime}\hat{C}_{\ell^{\mathstrut}_3,\nu^{\mathstrut}_3\nu_3^\prime} 
\right]\;,
\end{equation}
where $\Delta(\ell_1,\ell_2,\ell_3)$ counts the number of triangular configurations. Note that most of the signal arises from configurations where $\ell_1\neq\ell_2\neq\ell_3$, for which $\Delta = 1$.

The SNR for the bispectrum can be calculated as
\begin{equation}
\Sigma^2(\leq \ell_\mathrm{max}) = \sum_{\ell_\mathrm{min}\leq\ell_1\leq\ell_2\leq\ell_3}^{\ell_{\mathrm{max}}} \bm{B}^T(\ell_1,\ell_2,\ell_3) \bm{C_B}^{-1}(\ell_1,\ell_2,\ell_3)\bm{B}(\ell_1,\ell_2,\ell_3)\;,
\end{equation}
where, again, we bundled all the bispectra $B_{\nu_1,\nu_2,\nu_3}(\ell_1,\ell_2,\ell_3)$ at a single multipole combination into the vector $\bm{B}(\ell_1,\ell_2,\ell_3)$, with the ordering $\nu_1\leq\nu_2\leq\nu_3$. The covariance matrix $\bm{C_B}$ is the bundled version of Eq. (\ref{eq:covariance_bispectrum}). As a result, the Fisher matrix assumes the following form:
\begin{equation}\label{eq:fisher_matrix_bispectrum}
F_{ij}(\bm{\theta}_0) =  \sum_{\ell_\mathrm{min}\leq\ell_1\leq\ell_2\leq\ell_3}^{\ell_{\mathrm{max}}}\partial_i\bm{B}^T(\ell_1,\ell_2,\ell_3) \bm{C_B}^{-1}(\ell_1,\ell_2,\ell_3)\partial_j\bm{B}(\ell_1,\ell_2,\ell_3)\big|_{\bm{\theta} = \bm{\theta}_0} \;.
\end{equation}
For sake of computational tractability, we will bin the summation in the outer two sums over the $\ell$-modes and only apply the full sum for $\ell_3$ to take into account the correct behaviour of the Wigner 3$j$ symbol.

\subsection{Experimental setting and foreground modelling}
The choice of frequency bands, along with the white noise level and the resolution, are all listed in \autoref{table:experiments}. For the sky fraction, we assume $f_\mathrm{sky} = 0.6$ which will be used as the default value from now on unless stated otherwise.

The most challenging step in reconstructing maps of the CIB is the removal of contaminating signals such as the CMB or Galactic dust 
\citep{planck_collaboration_planck_2013,planck_collaboration_planck_2016-4,lenz_large-scale_2019}. The CMB signal can be extracted easily owing to its black-body nature, provided that the frequency coverage is sufficient. We will thus assume that the CMB has already been removed from the maps.
For Galactic dust emission, the situation is much more involved. In principle there are two approaches to deal with foreground contaminants: (i) include the dust model in the likelihood analysis; or (ii) remove the dust from the CIB maps. For the second case, \citet{tucci_cosmic_2016} used a method very similar to the ones used for CMB reconstruction \citep{tucci_limits_2005,stivoli_maximum_2010,stompor_maximum_2009,errard_robust_2016}. They showed that high frequencies are of paramount importance for a successful reconstruction of CIB maps. The reason for this is that the CIB and Galactic dust SEDs, Eq. (\ref{eq:galaxy_sed}) and (\ref{eq:sed_dust}), have very similar shape and differ only at higher frequencies. For this kind of CIB reconstruction, the noise variance of the CIB maps is given by
\begin{equation}\label{eq:reconstr_noise}
\Sigma^2_\mathrm{CIB}(\nu) =\frac{\sum_i \Theta^2_\mathrm{d}(i)/\sigma_i^2}{\mathrm{det}(\boldsymbol{A}^T\boldsymbol{N}^{-1}\boldsymbol{A})}\;,
\end{equation}
where $\boldsymbol{A}$ is the mixing matrix which describes how the two components (CIB and Galactic dust) mix in different frequency bands.  $\boldsymbol{N}$ is the noise covariance matrix. The reconstruction strategy then works as follows: Given a survey with $N_\nu$ frequencies, $N_\nu/2$ are used for the reconstruction, while the remaining channels are used for the CIB measurement. The reconstruction noise is then given by Eq. (\ref{eq:reconstr_noise}) which adds to the observed spectrum. In reality, there may be dust residuals in the CIB maps after the reconstruction. Furthermore, the reconstruction may also remove signal from the CIB itself owing to the very similar shape of the SEDs. We will ignore the latter effect in the following, but one should bear in mind that this could be an important source of systematics.

The signal-to-noise ratio for the power spectrum and the bispectrum is shown in the left and right panel of \autoref{fig:SNR_ideal}, respectively. We assume a fully reconstructed CIB in $8$ frequency bands. This means that we used another $8$ frequency bands, which are not shown in the analysis, to remove the galactic dust. The resulting reconstruction noise is assumed to be subdominant on the scales shown in the plot.
Different colours indicate the amount of residual Galactic dust contribution in the power spectrum and bispectrum. 
In the left panel, the solid lines show the cumulative signal-to-noise ratio, while the dashed lines correspond to the signal-to-noise ratio contribution at each multipole. The position of the kink in the dashed curves corresponds to the angular resolution of \texttt{LiteBird}. It should be noted that the signal-to-noise ratio is still rising at $\ell > 10^3$. However, as seen in \autoref{fig:Planck_spectra}, the spectra are dominated by the shot-noise contribution which itself carries very few information about cosmology, and is mainly sourced by the scatter in the $M-L$ relation.
The right panel shows only the cumulative signal-to-noise ratio for the bispectrum. Clearly it is dominated by the noise on angular scales smaller than the power spectrum. Furthermore, the total CIB bispectrum signal strength is approx. 4 times smaller than that of the power spectrum. 

The second possibility is to fit all the components separately. In particular we can write the observed signal as 
\begin{equation}\label{eq:splitting_real_sapce}
\delta I_\nu(\boldsymbol{\hat n}) = \delta I^\mathrm{CIB}_\nu(\boldsymbol{\hat n}) + \delta I^\mathrm{SN}_\nu(\boldsymbol{\hat n}) + \delta I^\mathrm{d}_\nu(\boldsymbol{\hat n})\;,    
\end{equation}
for CIB clustering, shot noise and the dust component respectively. In harmonic space we obtain a similar splitting on the power spectra level since the different signals are spatially uncorrelated:
\begin{equation}\label{eq:splitting_power}
\boldsymbol{\hat{C}}(\ell) = \boldsymbol{C}_\mathrm{CIB}(\ell) + \boldsymbol{N}(\ell) +  \boldsymbol{C}_\mathrm{d}(\ell)\;,
\end{equation}
where the first to terms are given by Eq. (\ref{eq:observed_spectra}) plus the shot noise contribution and the last term by Eq (\ref{eq:dust_power_amplitude}). A similar equation can be found for higher order spectra.

\section{Results}
\label{sec:results}
In this section we discuss the constraining power of CIB measurements on cosmological and HOD parameters in the absence and presence of foregrounds. The power spectrum and bispectrum analysis is discussed separately. Finally we also describe the combination of power and bispectrum. Throughout this section, we use the experimental settings specified in \autoref{table:experiments} with a sky-fraction $f_\mathrm{sky} = 0.6$.

\subsection{Power spectrum}
Firstly, we are interested in the sensitivity of the experiment described in \autoref{table:experiments} for the case where we fit all components simultaneously as outlined in the previous section. To this end, we fit all the components at the power spectrum and bispectrum level, rather than at the map level. In particular, we fix the slope $\alpha_\mathrm{d}$ and fit for a free dust amplitude at each frequency. We thus allow for slightly more flexibility in the SED modelling and, at the same time, ensure that the correlations are still described by (\ref{eq:frequency_correlation}). For the dust component, we therefore have $N_\nu$ free parameters. The shot-noise is fitted in each of the $N_\nu(N_\nu+1)/2$ pairs of frequency band separately, subject to satisfy the Cauchy-Schwarz inequality. This is very similar to the procedure outlined in \citet{feng_multi-component_2018}. This amounts to $N_\nu(N_\nu+1)/2$ additional parameters. The clustering signal of the CIB is fitted by varying both the CIB and cosmological parameters. This includes the total mass $\sum m_\nu$ of neutrinos, which reduce the small-scale clustering amplitude. We have consistently taken into account the impact of the resulting scale-dependent growth \citep{bond_massive_1980} on the linear power spectrum and on the halo mass function through the variance, Eq. (\ref{eq:variance_scale} \citep{saito_nonlinear_2009,ichiki_impact_2012,castorina_cosmology_2014}. In particular the cold dark matter density gets reduced to
\begin{equation}
\Omega_{\mathrm{cdm}} \equiv \Omega_{\mathrm{m}} - \Omega_{\rm b} - \frac{\sum m_{\nu}}{ 93.14 h^{2}}\;,
\end{equation}
with the mass of the neutrinos in eV.
\begin{figure}
\begin{center}
\includegraphics[width = 0.45\textwidth]{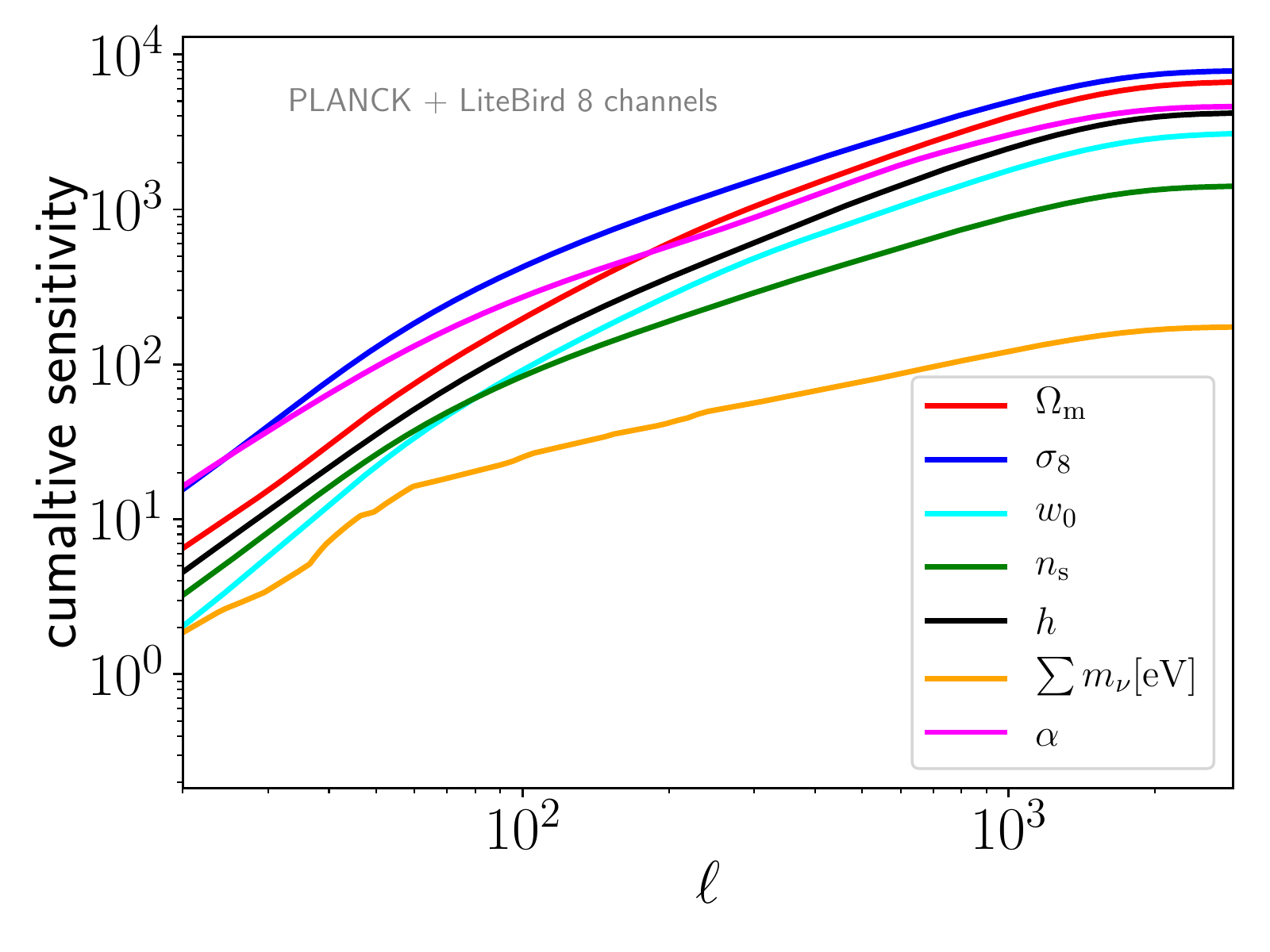}
\includegraphics[width = 0.45\textwidth]{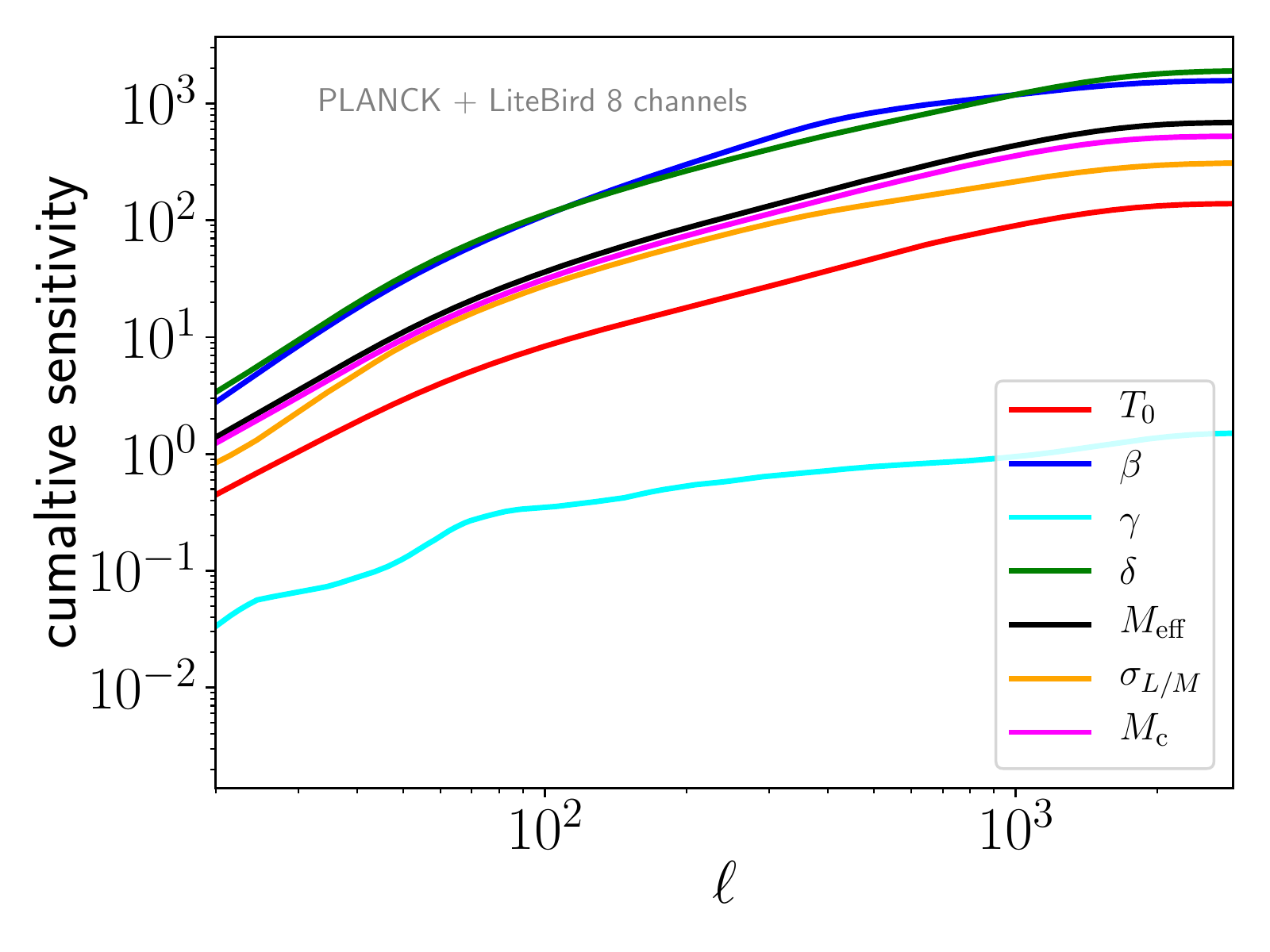}
\caption{Cumulative sensitivity of the power spectrum marginalized over the shot noise and dust  contributions as a function of multipole $\ell$. That is, we sum up Eq. (\ref{eq:fisher_matrix_gaussian_power}) up to multipole $\ell$. The Fisher matrix is still conditionalized on  the cosmological, HOD and CIB parameters, thus they are still fixed to their respective fiducial values. The experimental settings are summarized in \autoref{table:experiments}. \textit{Left}: sensitivity om cosmological parameter \textit{Right}: sensitivity on HOD and CIB parameters.}
\label{fig:sensitivity_power}
\end{center}
\end{figure}
In \autoref{fig:sensitivity_power}, we show the cumulative sensitivity (i.e. the Fisher information), Eq.~(\ref{eq:fisher_matrix_gaussian_power}), up to multipole $\ell$ marginalized over the shot noise and dust parameters. The sensitivity saturates above $\ell\approx 10^3$ due to the shot noise being the dominating contribution. Since the shot noise merely acts as a nuisance parameter in our model, we will restrict our analysis to $\ell \leq 10^3$ from now on. Depending on the focus of the analysis however, digging into the shot noise at higher multipoles may provide additional information about the physical modeling of the CIB \citep{shang_improved_2012,penin_non-gaussianity_2014,lacasa_non-gaussianity_2014}. 
For the minimum multipole we choose a conservative value of $\ell_\mathrm{min} = 50$ to reduce the contamination by foregrounds.

\begin{figure}
\begin{center}
\includegraphics[width = 0.9\textwidth]{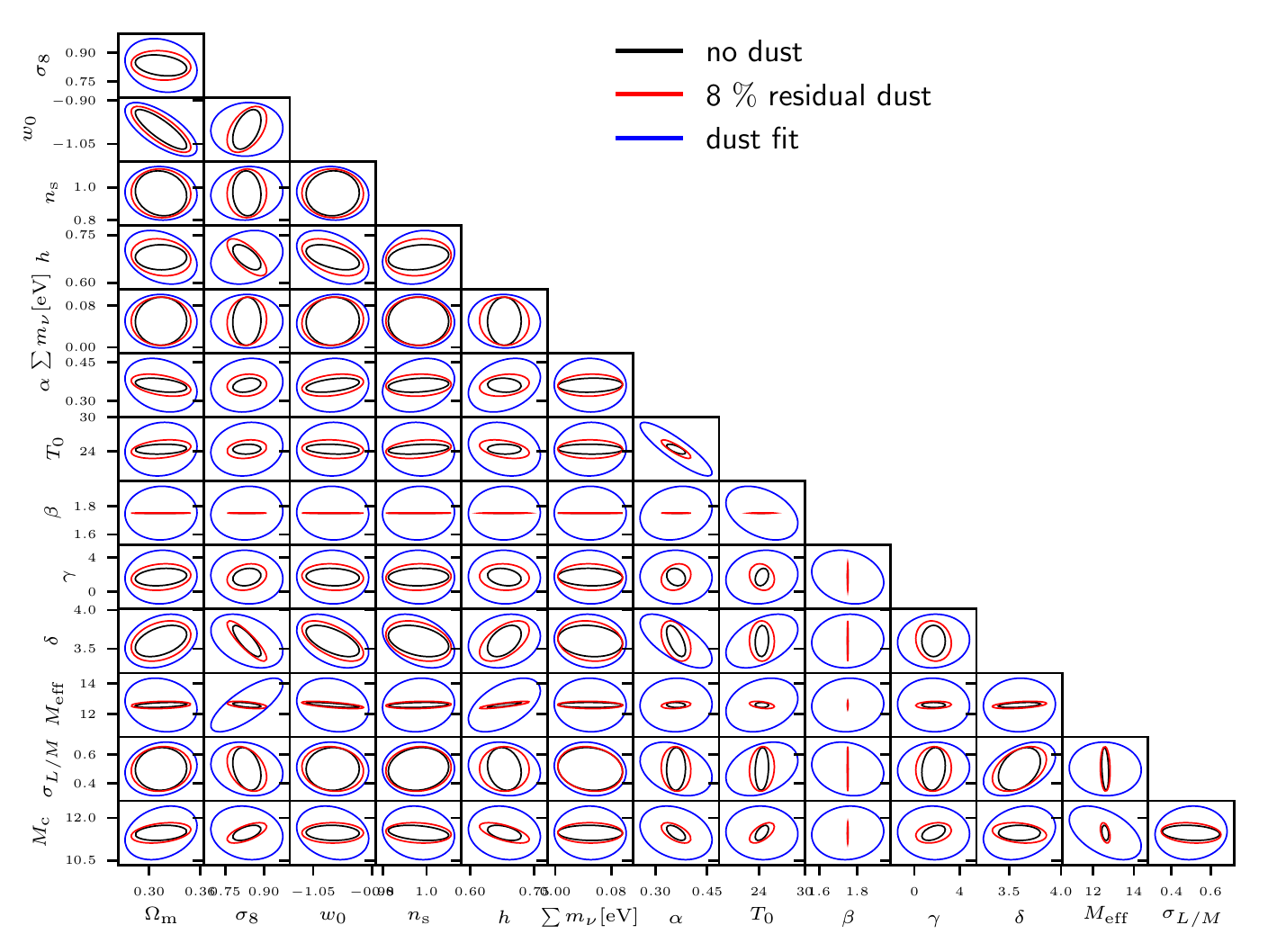}
\caption{$1\sigma$ constraints on CIB and cosmological parameters for the survey described in \autoref{table:experiments} using the power spectrum only with a sky fraction of 60 percent. The multipole range considered is $\ell\in[50,1000]$. The black ellipse corresponds to an ideal CIB survey without any dust residuals, while the red ellipse has eight percent dust residual at the power spectrum level. Blue ellipses fit the dust component and marginalize over it.}
\label{fig:fisher_dust_power}
\end{center}
\end{figure}

\autoref{fig:fisher_dust_power} shows a triangle plot with the $1\sigma$ contours. Only measurements at multipoles $50<\ell < 1000$ have been considered, and the general settings described in \autoref{table:experiments} has been used. The black and red ellipse show the constraints when the dust has been completely removed, or cleaned at the $92\% $ level, respectively. Blue ellipses correspond to the case where the dust and shot-noise components have been marginalized over. Clearly, the impact of the dust residuals and the $N_\nu(N_\nu+1)/2$ additional shot noise amplitudes strongly reduces the possible constraints on the CIB and HOD parameters. Interestingly, the cosmological parameters are largely unaffected. This result depends slightly on the fact that we assume a fixed power law for the spatial correlations of the foreground dust. However, even if this assumption is relaxed, the overall loss in precision is rather small. The amplitude of the power spectrum, $\sigma_8$, experiences the largest loss in precision.  For the CIB parameters, the biggest effect can be seen for $\beta$. This can be understood from the fact that $\beta$ is strongly degenerated with the dust, since it describes the modification to the black-body spectrum of the CIB and, therefore, can be constrained precisely with aid of different frequency channels and their cross-correlations.
One would, in principle, expect a similar effect for the high frequency power-law slope of the CIB's SED. However, most of the frequencies considered here lie below the peak of the SED across most of the redshift range probe by the CIB. This is the reason why it remains largely unconstrained.
Overall, large effects can be seen for all parameters associated with the infrared luminosity of the galaxies (\ref{eq:luminosity_mass_relation}) with the exception of $\sigma_{L/M}$.  One could also allow for the amplitude of the CIB power spectrum to vary with frequency. This would increase the errors on the SED parameters even further. However, the constraints on the cosmological parameters would be unchanged. 

\subsection{Bispectrum}
\begin{figure}
\begin{center}
\includegraphics[width = 0.9\textwidth]{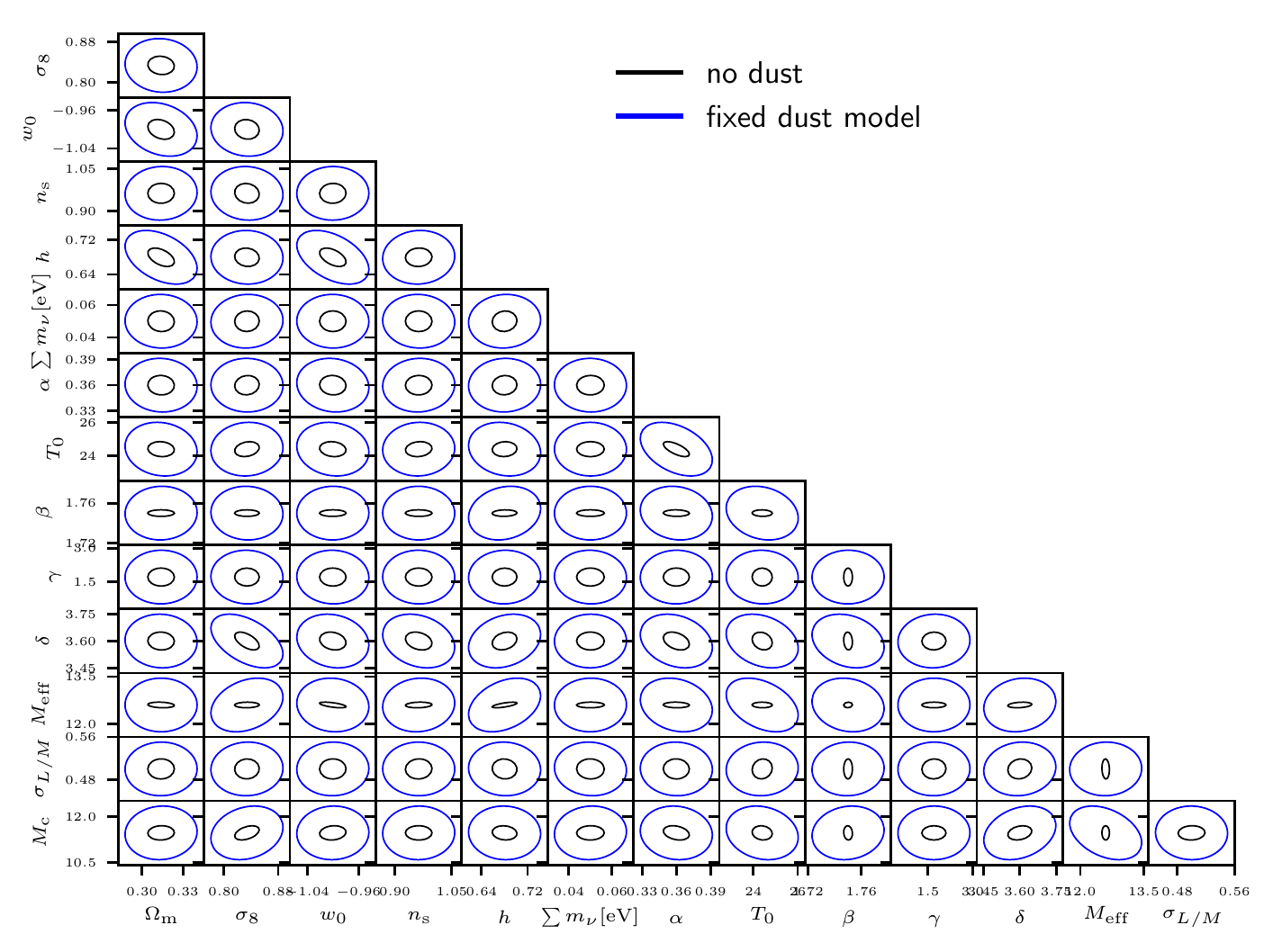}
\caption{Marginal contours for cosmological, CIB and HOD parameters using the survey settings described in \autoref{table:experiments}. Red ellipses show the constraints from the bispectrum alone while blue ellipses correspond to the dashed black contours in \autoref{fig:fisher_dust_power}. Black ellipses show the constraints from the combination of the power spectrum and the bispectrum analysis. The multipole range considered is $\ell\in[50,1000]$ and the sky fraction is 60 percent.}
\label{fig:fisher_dust_bispectrum}
\end{center}
\end{figure}
Before we present the results for the Fisher analysis of the bispectrum, we stress again that we do not consider contributions to the covariance from the $3-$, $4-$ or $6-$point correlation functions. Consequently, we likely underestimate the actual noise level. Nonetheless, note that the term $C_\ell^3$ dominates the terms stemming from $B_\ell^2$ at most angular scales. The individual contributions to the covariance Eq. (\ref{eq:covariance_bispectrum}) are, again, given by the cosmic variance of the CIB, Galactic dust, shot noise and instrumental noise.
Although the bispectrum is measured with less significance than the power spectrum, its sensitivity to non-linear parameters can be much higher, resulting into tighter constraints. Moreover, parameter degeneracies can be very different compared to an analysis of the power spectrum solely. 

\autoref{fig:fisher_dust_bispectrum} shows the 1$\sigma$ error contours for a bispectrum analysis in which Galactic dust has been fully removed (black), and in which the dust is still present as noise (black). 
Comparing this to the results obtained for the power spectrum, we see that any residual dust affects most parameters equally, with $\beta$ and $M_\mathrm{eff}$ being exceptions. The reason is twofold: the overall noise in the bispectrum analysis is higher, and most of the signal originates from smaller angular scales where the dust dominates (compare the panels of \autoref{fig:SNR_ideal}).

\begin{figure}
\begin{center}
\includegraphics[width = 0.9\textwidth]{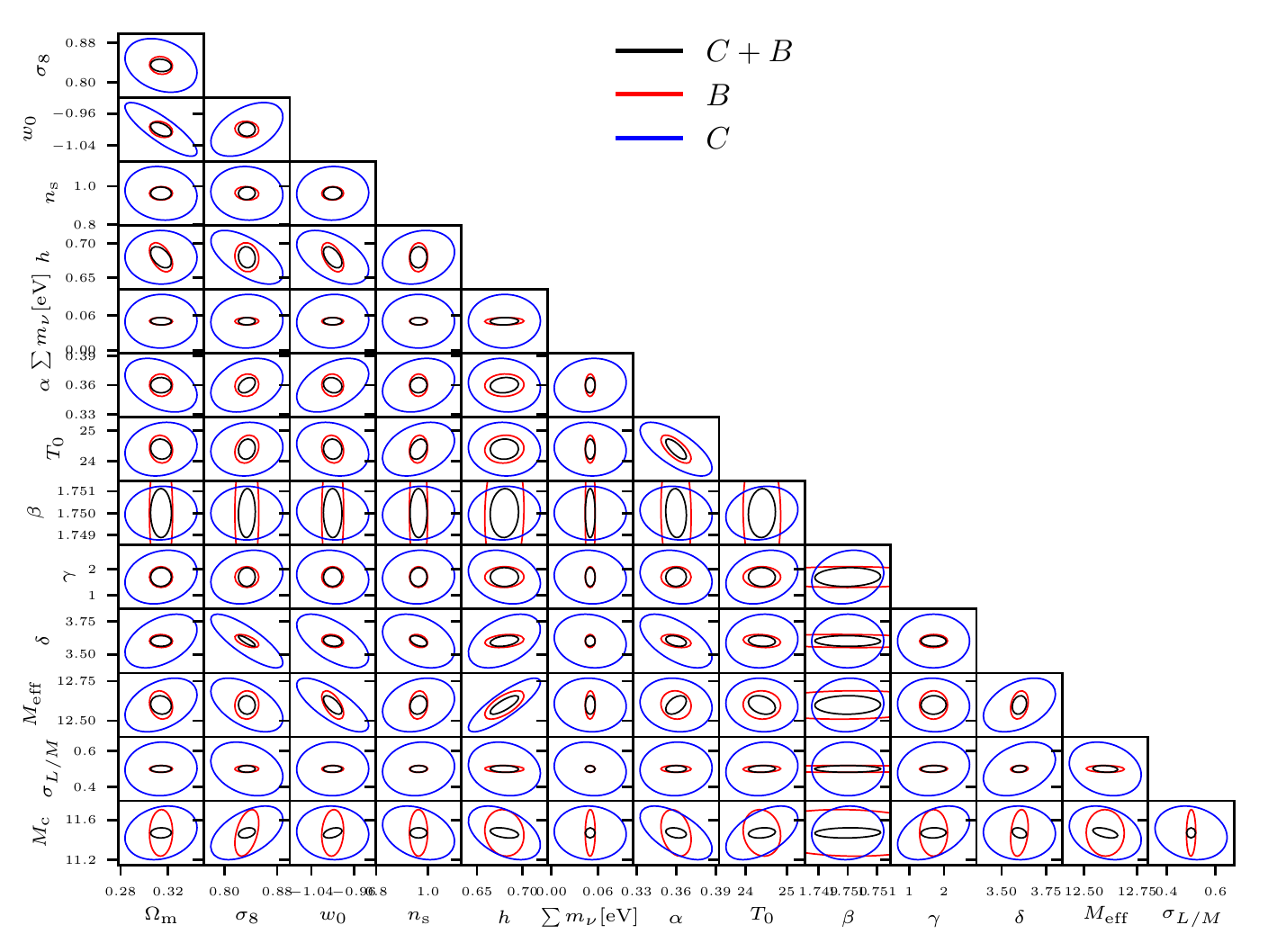}
\caption{Marginal contours for cosmological, CIB and HOD parameters using the survey settings described in \autoref{table:experiments}. Red ellipses show the constraints from the bispectrum alone while blue ellipses correspond to the dashed black contours in \autoref{fig:fisher_dust_power}. Black ellipses show the constraints from the combination of the power spectrum and the bispectrum analysis. The multipole range considered is $\ell\in[50,1000]$ and the sky fraction is 60 percent.}
\label{fig:fisher_no_dust_bispectrum_power}
\end{center}
\end{figure}

In order to compare the power spectrum and the bispectrum analysis, we show in \autoref{fig:fisher_no_dust_bispectrum_power} the 1$\sigma$ contours for both analysis when the CIB maps are assumed to be dust free. Furthermore, the result of a combined analysis is shown in black. For the latter, the bispectrum and power spectrum have been treated as two independent probes. Since the covariance between the two probes has no Gaussian contribution, the auto-correlations in the covariance are expected to dominate the cross-correlations, which justifies our assumptions. The marginal constraints of this figure are summarized in \autoref{table:constraints}. Clearly, the power spectrum is outperformed by the bispectrum for the cosmological parameters, usually yielding a factor $3-4$ improvement. The biggest improvement arises for the sum of the neutrino masses.
Interestingly, we find that the degeneracy directions are quite similar for the cosmological parameters. Consequently, the combination of power spectrum and bispectrum does not yield a substantial improvement for these parameters. The situation is very similar for the HOD and CIB parameters. However, it is possible to break degeneracies including $M_\mathrm{c}$ which yields  much tighter constraints for $M_\mathrm{c}$ when combining both probes (cf. \autoref{table:constraints}). 

\begin{table}
\begin{center}
\begin{tabular}{c|ccc}
parameter & $\sigma_C$ &$\sigma_B $ &  $\sigma_{C+B}$
\\ \hline\hline

$\Omega_\mathrm{m}$  & $   0.0201 \; ( 6.39 )$ & $   0.0064 \; ( 2.05 )$ & $   0.0058 \; ( 1.85 )$ \\
$\sigma_8$  & $   0.0361 \; ( 4.33 )$ & $   0.0120 \; ( 1.44 )$ & $   0.0084 \; ( 1.01 )$ \\
$w_0$  & $   0.0449 \; ( 4.49 )$ & $   0.0136 \; ( 1.36 )$ & $   0.0116 \; ( 1.16 )$ \\
$n_\mathrm{s}$  & $   0.0919 \; ( 9.55 )$ & $   0.0232 \; ( 2.41 )$ & $   0.0217 \; ( 2.25 )$ \\
$h$  & $   0.0262 \; ( 3.85 )$ & $   0.0142 \; ( 2.09 )$ & $   0.0103 \; ( 1.52 )$ \\
$\sum m_\nu[\mathrm{eV}]$  & $   0.0307 \; ( 61.39 )$ & $   0.0042 \; ( 8.47 )$ & $   0.0042 \; ( 8.38 )$ \\
$\alpha$  & $   0.0181 \; ( 5.02 )$ & $   0.0076 \; ( 2.10 )$ & $   0.0052 \; ( 1.44 )$ \\
$T_0$  & $   0.5771 \; ( 2.37 )$ & $   0.3007 \; ( 1.23 )$ & $   0.2169 \; ( 0.89 )$ \\
$\beta$  & $   0.0008 \; ( 0.05 )$ & $   0.0022 \; ( 0.12 )$ & $   0.0007 \; ( 0.04 )$ \\
$\gamma$  & $   0.6877 \; ( 40.45 )$ & $   0.2671 \; ( 15.71 )$ & $   0.2422 \; ( 14.25 )$ \\
$\delta$  & $   0.1337 \; ( 3.71 )$ & $   0.0327 \; ( 0.91 )$ & $   0.0267 \; ( 0.74 )$ \\
$M_\mathrm{eff}$  & $   0.1122 \; ( 0.89 )$ & $   0.0593 \; ( 0.47 )$ & $   0.0387 \; ( 0.31 )$ \\
$\sigma_{L/M}$  & $   0.0972 \; ( 19.44 )$ & $   0.0125 \; ( 2.50 )$ & $   0.0123 \; ( 2.47 )$ \\
$M_\mathrm{c}$  & $   0.1771 \; ( 1.54 )$ & $   0.1537 \; ( 1.34 )$ & $   0.0336 \; ( 0.29 )$ \\

\end{tabular}
\caption{Marginal constraints for cosmological, CIB and HOD parameters using the survey settings described in \autoref{table:experiments}. The first two columns display the constraints obtained from the power spectrum and the bispectrum, respectively. The third column summarizes the percental change of the constraints. The last columns gives the error achievable with a joint analysis of the CIB power spectrum and bispectrum. For all absolute errors, the relative error is shown in percent in brackets.}
\label{table:constraints}
\end{center}
\end{table}

\section{Conclusions}
\label{sec:conclusions}
In this paper, we investigated the information content of CIB anisotropies using their power and bispectrum. Previous work \citep{penin_non-gaussianity_2014} mainly focused on the impact of the HOD parameters, mass function and galaxy formation on the CIB's bispectrum. Our approach is complementary since we investigate the information content of the CIB also with respect to the cosmological parameters, exploiting multiple frequency bands and their cross-correlations.

We analytically modelled the anisotropies using the halo model and the approach introduced in \citet{shang_improved_2012}. 
In particular, the model assumes that the clustering of halos on large scales is reasonably described by the combination of the halo model with a HOD. Each galaxy is then assigned a specific IR emissivity, which is fully specified by the SED and by a mean mass-luminosity relation. We did not explicitly model the shot noise contribution, which can be done with empirical models \citep{bethermin_modeling_2011,wu_minimal_2017}. Overall, our phenomenological model is sufficient for the multipole range considered here, although it would be desirable to phrase it as a rigorous bias expansion \citep{desjacques_large-scale_2018}.

The theoretical predictions were applied to forecast constraints on HOD, CIB and cosmological parameters for a combined survey of \texttt{Planck} and \texttt{LiteBird} with a total of eight frequency channels between 200 and 900$\;\mathrm{GHz}$, using all the auto- and cross-correlations available.
Furthermore, we studied the impact of Galactic dust emission which we assumed to be strongly correlated over the relevant range of frequencies, and whose angular power spectrum was modelled as a power law. In particular, we investigated the impact of dust residuals on the constraints yielded by the power spectrum. Furthermore, we explore the sensitivity of the bispectrum, and of its combination with the power spectrum, to the model parameters. We summarize our main results as follows:

(i) For the experiments considered here, the power spectrum CIB signal of the clustering component can be measured by a few hundred $\sigma$ when the foreground dust is at least partially removed (with a maximum of eight per cent dust residuals). For the bispectrum the SNR is roughly four times smaller.

(ii) Confidence intervals on cosmological parameters are not strongly affected by residual dust in the maps. Even if the dust model and the shot noise are treated as free parameters, the cosmological parameters are still constrained down to an uncertainty of $\sim 10\%$ even after marginalization.

(iii) The clustering components (i.e. all at least partially connected parts of the correlation functions) of the bispectrum suffer more strongly from residual dust, since the shot noise component becomes important at lower multipoles, where the dust contribution is more dominant. 

(iv) Overall the power spectrum yields weaker constraints (by a factor of four) than the bispectrum for almost all parameters of the model -- assuming that both the power spectrum and bispectrum model are equally accurate over the multipole range considered. Degeneracy directions are very similar between the power spectrum and the bispectrum analysis. Therefore, the combination of both statistics yields substantial improvement for the HOD parameters solely. However, we caution that this might be due to the simplified bias description generally adopted in such halo model approaches.

We plan to refine the halo model description of the CIB by including other relevant terms from the bias expansion and, possibly, taking into account scatter in the HOD parameters to model better the shot noise. Further analysis could include a study of the cross-correlation between CIB and LSS probes -- such as spectroscopic galaxy surveys like \texttt{SPHEREx} \citep{dore_cosmology_2014} -- that can probe a similar redshift range; and an application of CIB bispectrum measurements to constrain primordial non-Gaussianities. 
%\bsp

\section*{Acknowledgments}
R.R. and V.D. acknowledges support by the Israel Science Foundation (grant no. 1395/16). R.R and S.Z. furthermore acknowledge support by the Israel Science Foundation (grant no. 255/18).

\label{lastpage}
\bibliographystyle{mnras}
\bibliography{My_Library}
\end{document}